\documentclass[prb,twocolumn,showpacs,amsmath,amssymb]{revtex4}
\usepackage{graphicx}
\usepackage{dcolumn}
\usepackage{bm}
\usepackage{epsfig}

\begin{document}


\title{Magnetism in the $S=1$ frustrated antiferromagnet
GeNi$_{2}$O$_{4}$ studied using implanted muons}

\author{T. Lancaster}
\email{t.lancaster1@physics.ox.ac.uk}
\author{S. J. Blundell}
\author{D. Prabhakaran}
\author{P. J. Baker}
\author{W. Hayes}
\affiliation{
Clarendon Laboratory, Oxford University Department of Physics, Parks
Road, Oxford, OX1 3PU, UK
}
\author{F. L. Pratt}
\affiliation{
ISIS Facility, Rutherford Appleton Laboratory, Chilton, Oxfordshire OX11 0QX, UK}

\date{\today}

\begin{abstract}
We present the results of a muon-spin relaxation
study of GeNi$_{2}$O$_{4}$. We provide further clarification
of  the two transitions to the antiferromagnetic state
and measure the magnetic field dependence of the heat capacity
up to 14~T. Both oscillatory and relaxing signals are observed 
below the lower transition (at temperature $T_{\mathrm{N2}}$) in the 
muon-decay positron asymmetry spectra, arising 
from two distinct types of magnetic environment. A possible
explanation is suggested in terms of  the separate ordering of
two magnetic subsystems, one of which does not order fully down
to the lowest measured temperature. 
\end{abstract}

\maketitle

\section{Introduction}

Materials experiencing geometric frustration
have been the topic of much recent interest. 
Their rich behavior is
due to a large ground-state degeneracy,
which renders them highly unstable to perturbations
\cite{moessner}.
In the case of a frustrated model system comprising
Heisenberg-type antiferromagnetic nearest-neighbor interactions, 
cooperative paramagnetic
behavior (i.e.\ fluctuations at all temperatures down to absolute zero) is
expected. This state of affairs is altered, however, by perturbations due
to other interactions (including next-nearest neighbor, dipole, crystal
field etc.) which cause a variety of low temperature states to be
realized, including spin-liquids \cite{gardner} and
spin-ices \cite{bramwell} in some rare-earth pyrochlores and structural
phase transitions for some spinels with transition metal ions on the frustrated B
sublattice \cite{lee,huang}.

Many geometrically frustrated 
systems with half-integer spins relieve the frustration by undergoing
a structural phase transition at low temperatures to a magnetically
ordered state; this is the case for the spinels
ZnCr$_{2}$O$_{4}$ ($S=3/2$) \cite{lee}, 
GeCo$_{2}$O$_{4}$ ($S=3/2$) 
and powder samples of ZnFe$_{2}$O$_{4}$ ($S=5/2$)
\cite{huang,schiessl}
(although we note that a magnetic transition is not observed
in single crystal samples \cite{usa}) .
The integer spin ($S=1$) frustrated antiferromagnet 
GeNi$_{2}$O$_{4}$ has therefore been of recent interest, 
since it 
undergoes a transition to an
antiferromagnetically ordered state
below $\sim 12$~K (observed in both powder and
single crystal samples \cite{bertaut,crawford,hara})
with no accompanying structural transition. 
GeNi$_{2}$O$_{4}$ has the normal spinel structure, having Ni$^{2+}$
ions (3d$^{8}$) at the vertices of corner sharing tetrahedra 
(the spinel B site), 
coordinated by a nearly regular octahedron of oxygen ions, resulting
in a $^{3}A_{2\mathrm{g}}$ triplet ground state. The crystal field 
lowers the degeneracy of this triplet, further splitting it into a
close lying spin-singlet and doublet expected to be separated by 
only a few cm$^{-1}$. 
Recent experimental studies of this system by Crawford {\it et al.\
}\cite{crawford} 
show that the ordered state is reached by
two separate transitions at $T_{\mathrm{N1}}=12.13$~K and 
$T_{\mathrm{N2}}=11.46$~K. 
It was found from heat capacity measurements  that
the magnetic entropy $S_{\mathrm{mag}}$
of GeNi$_{2}$O$_{4}$ is only half of the expected $2R \ln 3$ per mole 
\cite{crawford},
with the same measurements suggesting the existence of both gapped
and gapless excitations within the N\'{e}el state. 

\begin{figure}
\begin{center}
\epsfig{file=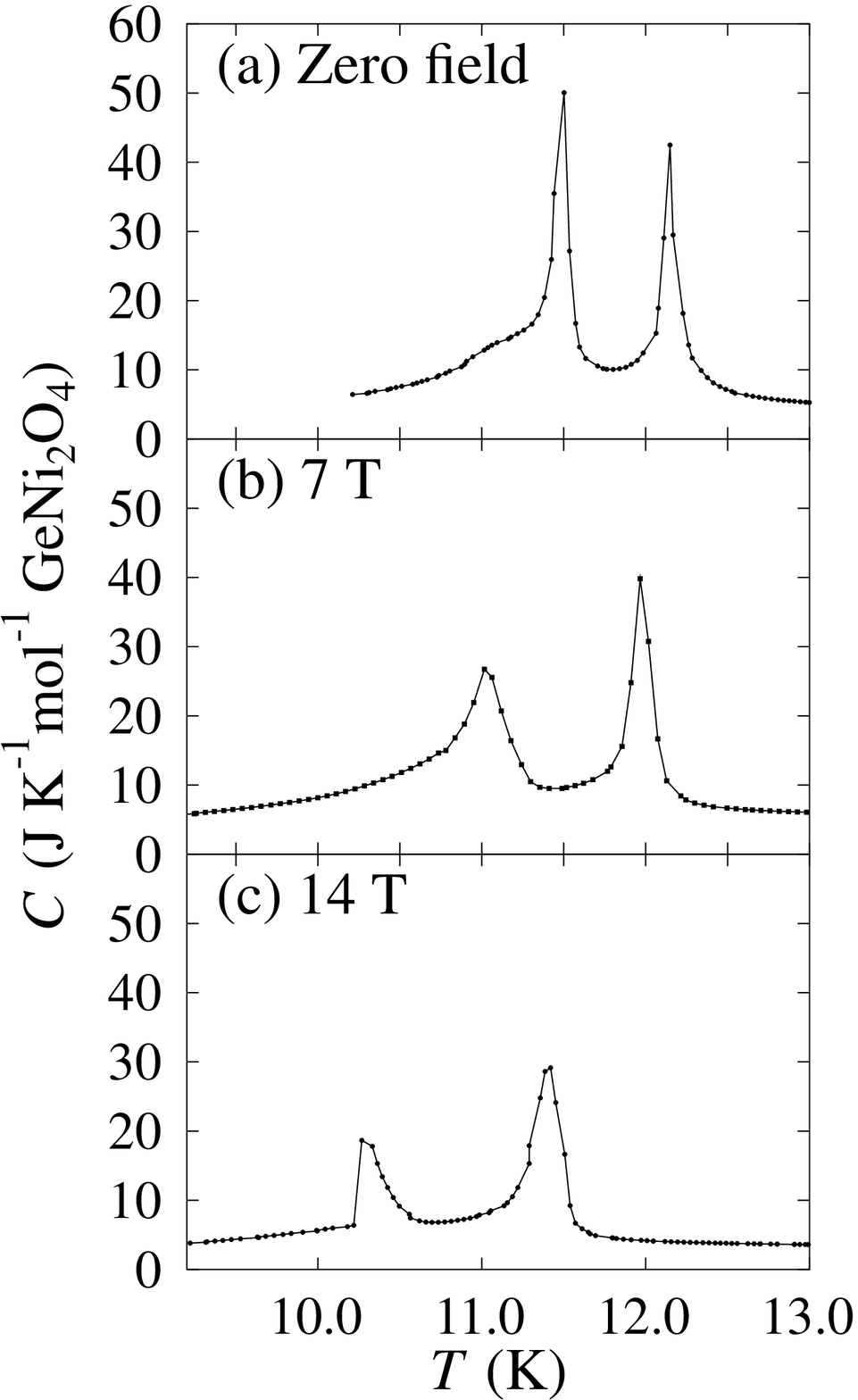,width=4.5cm}
\epsfig{file=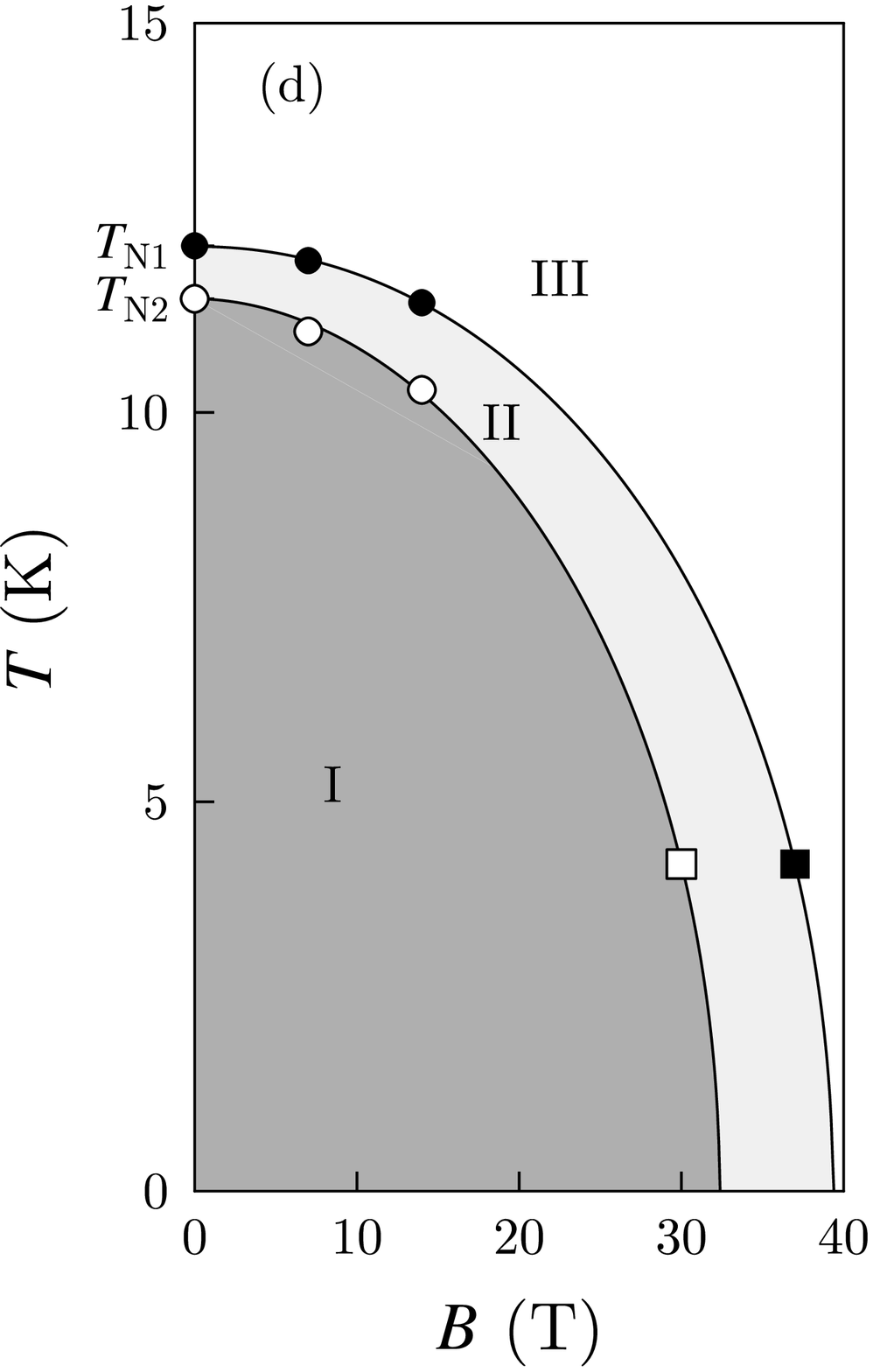,width=3.95cm}
\caption{Heat capacity measured in (a) zero field 
and in an applied magnetic field of (b) 7\,T and (c) 14\,T.
(d) Magnetic phase diagram deduced from (a)-(c) (circles)
and from spin flops observed in high field magnetization
measurements\cite{diaz} (squares) (phases I and II are
antiferromagnetic phases, phase III is paramagnetic). 
\label{hc}}
\end{center}
\end{figure}

\begin{figure*}
\begin{center}
\epsfig{file=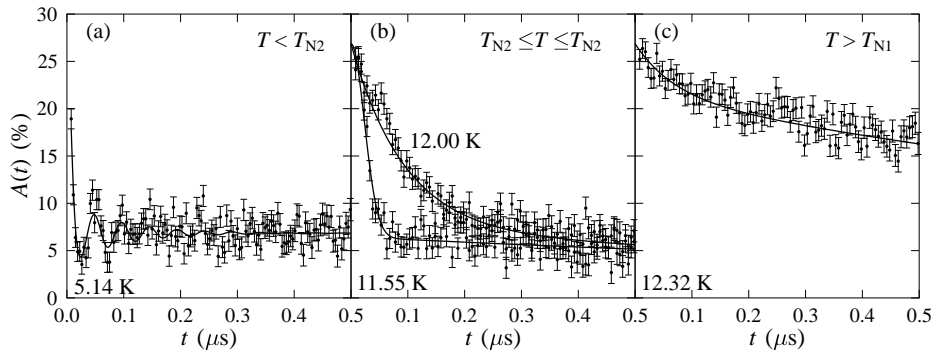,width=13cm}
\caption{ZF spectra measured 
(a) at 5.14~K, in phase I ($T < T_{\mathrm{N2}}$), 
(b) at 11.55~K and 12.00~K, in phase II
($T_{\mathrm{N2}} \leq T \leq T_{\mathrm{N1}}$) and
(c) at 12.32~K, in phase III ($T > T_{\mathrm{N1}}$). 
The oscillations seen in (a) are observed to vanish above
$T_{\mathrm{N2}}$ (b), with the Gaussian component vanishing
 above $T_{\mathrm{N1}}$ (c).\label{data}}
\end{center}
\end{figure*}

In recent years, muon-spin relaxation ($\mu^{+}$SR) measurements  have
been highly successful in observing frustration-related behavior in several
frustrated systems, including the observation of  spin-ice behavior
and cooperative paramagnetism (see e.g.\ Ref.~\onlinecite{steve2,gardner}).
In this paper we report on the first muon-spin relaxation
measurements on  GeNi$_{2}$O$_{4}$. We have probed the magnetically
ordered region from a local viewpoint and provide evidence for muons
stopping in separate magnetic environments in the ordered state. It is
suggested that the double transition may be due to the separate
ordering of these two subsystems. 

\section{Experimental}

Polycrystalline GeNi$_{2}$O$_{4}$ was prepared in a solid-state reaction
 using high purity ($>99.99$ \%) GeO$_{2}$ and NiO.  The
stoichiometric mixed powder was calcined in an O$_{2}$ flow atmosphere
at 1200$^{\circ}$C for 48 hours.  
The sample was characterized using X-ray 
diffraction analysis (which revealed no contribution from impurity phases) and
heat capacity measurements. 
The latter were carried out 
on a sintered pellet
sample on warming using a 14\,T Quantum Design PPMS system.
Heat capacity data taken in  zero field (ZF) and in an applied field of 
7\,T and 14\,T
are shown in Fig.\ref{hc}(a-c). The ZF results show two sharp 
maxima \cite{hcnote} at 
$T_{\rm N1}$ and $T_{\rm N2}$, in agreement
with previous studies \cite{crawford,hara}. In addition, we find
a small
shoulder below $T_{\mathrm{N1}}$, also observed by Hara {\it et al.}
\cite{hara} who speculated about
a possible third transition. 
The temperature of the two main peaks decreases markedly with applied
field (as expected for an antiferromagnet, see e.g.\ Ref.~\onlinecite{leidl}),
as shown in the phase diagram in Fig.~\ref{hc}(d). 
Both main peaks broaden and decrease in intensity with increasing field, 
the effect
being more pronounced for the lower temperature peak.
Also included
in Fig.~\ref{hc}(d)
are the two spin-flop transitions measured at 4.2~K in a recent high field 
magnetization study \cite{diaz}, 
and our phase diagram associates these with our
heat capacity peaks.

Zero-field muon-spin relaxation measurements have been made 
on polycrystalline GeNi$_{2}$O$_{4}$ using the DOLLY instrument at the
Swiss Muon Source, 
Paul Scherrer Institute (PSI), Villigen, Switzerland\cite{note}. 
The sample was wrapped in 25~$\mu$m Ag foil and mounted
on a Ag backing plate. 
In a $\mu^{+}$SR experiment, spin-polarized
positive muons are stopped in a target sample, where the muon usually
occupies an interstitial position in the crystal.
The observed property in the experiment is the time evolution of the
muon spin polarization, the behavior of which depends on the
local magnetic field $B$ at
the muon site, and which is proportional to the
positron asymmetry function \cite{steve} $A(t)$.

\section{Results}

Our $\mu^{+}$SR data allow us to identify three distinct regions
of temperature as follows:
(a) phase I ($T < T_{\mathrm{N2}}$), (b) 
phase II ($T_{\mathrm{N2}} \leq T \leq T_{\mathrm{N1}}$)
and (c) phase III ($T > T_{\mathrm{N1}}$). We now discuss
each of these regimes in turn.

\subsection{Phase I ($T < T_{\mathrm{N2}}$)}
For phase I (Fig.\ref{data}(a)),
oscillations in the asymmetry are clearly discernible, though
with a rather small amplitude, and a fast initial relaxation
of the muon polarization is visible at early times. 
The oscillations are characteristic of a quasistatic local
magnetic field at the muon site, which causes a coherent precession of 
the spins of 
those muons with a component of their spin 
polarization perpendicular to
this local field; their presence
provides
strong evidence for the existence of long range magnetic order (LRO)
in phase I, in agreement with previous neutron diffraction measurements
\cite{crawford}. 
The frequency of the oscillations is given by 
$\nu_{i}=\gamma_{\mu} B_{i}/2 \pi$, where
$\gamma_{\mu}$ is the muon gyromagnetic ratio 
($\equiv 2 \pi \times 135.5$~MHz T$^{-1}$), and $B_{i}$ is the local field at
the $i$th muon site. 
The existence of oscillations in the asymmetry 
in a polycrystalline sample is a
signature of a narrow distribution of the magnitudes of the static
local magnetic fields, 
at symmetry-related muon sites in the crystal,
associated with LRO. 
A broader distribution
of these fields leads to the occurrence of the Kubo-Toyabe (KT)
function\cite{hayano}, 
which is well approximated by a Gaussian function at early times 
with relaxation rate $\sigma^{2}=\gamma_{\mu}^{2} \Delta B^{2} /2$, where
$\Delta B^{2}= \langle (B- \langle B \rangle)^{2}\rangle$ is the
second moment of the local field distribution. The recovery of 
asymmetry at late times, characteristic of the
KT function, is often lost due to the presence of slow dynamics
in the local field distribution.
Because of the additional fast relaxation signal in the data,
a full description also requires an additional component. This is
well described over the measured temperature regime
by a Gaussian functional form
$A_{2}\exp(-\sigma^{2}t^{2})$ with a large relaxation rate (see below). 
The large magnitude of the
relaxation at very low temperatures makes it impossible to 
unambiguously assign it a Gaussian lineshape over the entire
temperature range. However, the continuous temperature evolution
of $\sigma$ and the line-shape seen at temperatures $T >10$~K makes
this a reasonable assumption.

The
spectra are found to be best fitted in the range $T<T_{\mathrm{N1}}$
with the resulting functional form
\begin{eqnarray}
A(t)  = A_{1}
\left\{ a_{1} + a_{2} \exp(-\lambda_{1} t) \cos(2 \pi \nu_{1} t)
 \right\} \nonumber \\
+A_{2}\exp(-\sigma^{2}t^{2}) + A_{\mathrm{bg}},
\label{phaseI}  \end{eqnarray}
where $A_{\mathrm{bg}}$ represents a constant background contribution from
those muons that stop in the sample holder or cryostat tail\cite{fits}. 
The factor multiplied by $A_{1}$ accounts for the
component of the spectra associated with LRO (normalized to 1), 
expected to be made up from
contributions from
 those muons with spin components
parallel to the local magnetic field $a_{1}$ 
and those with perpendicular components $a_{2}$. 
The value of $a_{1}$ consistent with the measured spectra
is much higher than expected ($a_{1}=0.59$ compared to the 
expected value of 1/3). This may be 
due to muons stopping at similar sites of LRO 
but whose transverse spin contributions are rapidly dephased
(although we note that this should lead to a missing fraction of
asymmetry, which we do not observe); an unambiguous
assignment of the amplitudes for the lowest measured
temperatures is hindered by the large value of the relaxation
rate $\sigma$ so the situation may be more complex than considered
here. 

The amplitude
$A_{1}$ is found to be approximately constant up to $T
\approx 9$~K 
(Fig.\ref{fit}(a)), where
it is seen to decrease as the transition at $T_{\mathrm{N2}}$ is
approached from 
below. This is
accompanied by a corresponding rise in the amplitude of the Gaussian
component $A_{2}$. 
The existence of two distinct components in the measured spectra
(i.e.\ Gaussian and oscillatory) provides evidence for two sets of
muon site (or {\it subsystem}) in the material, such that a localised 
muon experiences 
one of two distinct
magnetic environments. The first of these subsystems, with occupancy 
proportional
to the amplitude $A_{1}$, has a sufficiently narrow distribution
of quasistatic magnetic fields that oscillations are observable. The
other subsystem, with occupancy proportional to $A_{2}$, is
associated with a wider
distribution of fields (preventing oscillatons being
observable) and slow dynamics preventing
any recovery of the asymmetry at later times. 
 The $\mu^{+}$SR
data show that the fraction of
the sample associated with the Gaussian component gives rise
to a very different local magnetic environment to that causing
the fraction associated with amplitude $A_{1}$. 
Since the muon is a local probe, it is not possible to conclude
whether the two magnetic environments suggested by the data are
spatially separated or spatially 
coexisting. If these two subsystems were spatially
separated, then the exchange of amplitudes seen as $T_{\mathrm{N2}}$
is approached from below would probably be due to those regions
associated with Gaussian relaxation growing at the expense
of those sites related to the oscillations. If, in contrast, the two 
environments
are coexisting, then the changes in amplitude may be due to a
reordering of the magnetic moments as the temperature is increased.

The frequency of the oscillations (Fig.\ref{fit}(b)) varies little
below the transition temperature 
$T_{\mathrm{N2}}$, at which point $A_{1} \rightarrow 0$,
indicating first-order behavior, in agreement with the
heat capacity measurements \cite{crawford}.
This could suggest that the subsystem associated with
the oscillating component undergoes an ordering transition
at $T_{\mathrm{N2}}$.
The stability of the oscillation frequency across the temperature
range means that we see no effect due to depopulation
of the upper level of the split $^{3}$A$_{2\mathrm{g}}$ triplet state at low
temperatures.

\subsection{\it Phase II ($T_{\mathrm{N2}} \leq T \leq T_{\mathrm{N1}}$)}
In phase II (Fig.\ref{data}(b))the oscillations vanish and the spectra are
described by relaxing components only. 
The Gaussian component persists
into this regime, with no discernible discontinuity in $\sigma$ at
$T_{\mathrm{N2}}$. This suggests that this component arises from the
same mechanism responsible for the corresponding component in 
Eq.\ref{phaseI}. Also evident in the spectra measured in
phase II is an exponential component 
$A_{3} \exp (-\lambda_{3} t)$. An exponential function is 
often the result of fast, dynamic fluctuations in the local field, in
which case the relaxation rate 
$\lambda_{3} \propto \gamma_{\mu}^{2} \Delta B^{2} / \delta$, where
$\delta$ 
is the 
fluctuation rate \cite{hayano}.
This behavior points to dynamic
fluctuations of a paramagnetic subsystem, coexisting with
the subsystem associated with the Gaussian component seen in phase I.
The data are therefore described by the resulting relaxation function
\begin{equation}
A(t) = A_{2} \exp(-\sigma^{2}t^{2})  + A_{3} \exp (-\lambda_{3} t)
 + A_{\mathrm{bg}}\label{phaseII}.
\end{equation}

\begin{figure}
\begin{center}
\epsfig{file=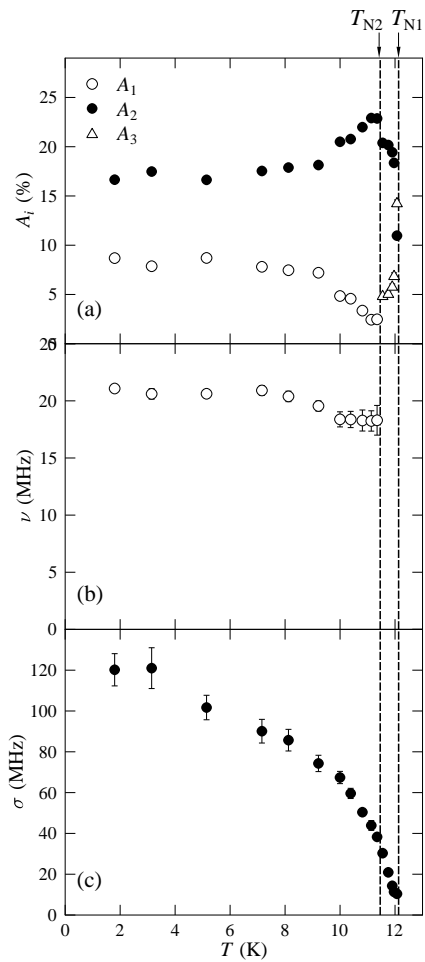,width=6.85cm}
\caption{Results of fits as described in the main text. 
(a) Amplitudes of the magnetic component $A_{1}$ (open circles) and 
Gaussian component $A_{2}$ (closed circles) as a function of
temperature. 
$A_{2}$ is seen
to rise as $T_{\mathrm{N2}}$ is approached from below. 
Between the transitions, $A_{2}$ decreases suddenly, while 
$A_{\mathrm{3}}$ increases (open
triangles). The error bars are smaller than the data points in this case. 
(b) The precession frequency remains approximately constant for
$1.8~\mathrm{K} \leq T \leq T_{\mathrm{N2}}$.
(c) The relaxation rate $\sigma$ is seen to fall smoothly across the
transition
at $T_{\mathrm{N2}}$.
\label{fit}}
\end{center}
\end{figure}

With increasing temperature, we see 
(Fig.\ref{fit}(a)) the increase in the amplitude of the fluctuating
component ($A_{3}$ increases) 
along with a decreasing fraction of the Gaussian component
($A_{2}$ decreases). 
The relaxation rate $\sigma$ decreases smoothly across the
entire temperature range (Fig.\ref{fit}(c)), 
showing no features corresponding to the transition at $T_{\mathrm{N2}}$
but vanishing at $T_{\mathrm{N1}}$. 
The smooth decrease of $\sigma$ with increasing temperature 
suggests that this parameter is dominated by the magnitude of the local
field distribution, as seen in
other muon studies of magnetically ordered systems \cite{tom}. This
may suggest that the subsystem associated with this component
undergoes an ordering transition at $T_{\mathrm{N1}}$. 

\subsection{{\it Phase III} ($T > T_{\mathrm{N1}}$)} 

In Phase III (Fig.\ref{data}(c)) the Gaussian component is absent and the
data are best described by a two exponential
form $A(t) = A_{3} \exp(-\lambda_{3} t) + A_{4}\exp(-\lambda_{4}
t)+A_{\mathrm{bg}}$,
typical of relaxation due to dynamic fluctuations, as would be 
expected in a purely paramagnetic material.

\section{Discussion}

The fact that only half of the expected magnetic entropy
$S_{\mathrm{mag}}=2 R \ln 3$ is
accounted for in heat capacity measurements made below 75~K
points to unconventional magnetic behavior in GeNi$_{2}$O$_{4}$.
The double transition is seen in neutron diffraction
measurements\cite{crawford} through
discontinuities in the intensity of the $\left( \frac{1}{2}, 
\frac{1}{2}, \frac{1}{2} \right)$
 magnetic Bragg
peak, the intensity of which is proportional
to the ordered volume-fraction multiplied by the magnetic
moment\cite{squires}. These
results suggest that LRO exists throughout the material with the
same wavevector at all temperatures below $T_{\mathrm{N1}}$. 

The muon provides us with a local perspective. Of the two magnetic
subsystems identified in the data, the one giving rise to
oscillations may unambiguously be attributed to the presence
of LRO. In phase I the magnetic moment (proportional
to the frequency $\nu_{1}$) shows little temperature dependence, in
contrast to the volume fraction, which decreases as $T$ tends to
$T_{\mathrm{N2}}$.
This component cannot be resolved in phase II, and the disappearance
of magnetic order in this subsystem may cause the discontinuity in the 
Bragg peak intensity seen in the neutron data. 

Since the neutron diffraction results strongly suggest that LRO
exists throughout the sample in phase II, it may be that the
Gaussian component associated with the other magnetic subsystem
seen in phases I and II could also be an artifact of
LRO. An incomplete ordering of the magnetic 
moments, for example, may 
provide sufficient spin disorder in this subsystem
to wash out an oscillatory
signal. The magnitude of the local magnetic
field in this subsystem may still be probed from the behavior of the relaxation
rate $\sigma$ (provided that any dynamics are slow and only weakly
temperature dependent), which decreases smoothly. The amplitude of this component
decreases with increasing temperature, before vanishing at
$T_{\mathrm{N1}}$, 
perhaps accounting for the
other discontinuity in the Bragg peak intensity. 

The observed behavior in
GeNi$_2$O$_4$ may, therefore, be analogous to Gd$_{2}$Ti$_{2}$O$_{7}$, 
in which there are two magnetic subsystems with different ordering
temperatures\cite{steward}. In that system ordering
takes place at the higher transition in one subsystem. At the
lower transition temperature the other subsystem weakly orders. 
If a similar mechanism were to be the cause of the observed behavior
in GeNi$_2$O$_4$, we note that the higher transition  at
$T_{\mathrm{N1}}$ would probably be
associated with a weakly ordering subsystem (with sufficient
spin disorder to prevent the observation of coherent oscillations)
followed by the ordering of the second subsystem below
$T_{\mathrm{N2}}$.

Such behavior in GeNi$_{2}$O$_{4}$ could also explain the missing
contribution to the magnetic entropy. 
In their heat capacity study Crawford {\it et al}.\ assume
that  $\lim_{T \rightarrow 0}\, S_{\mathrm{mag}}=0$ and detect
only 56.5~\% of the expected $S_{\mathrm{mag}}=2 R \ln 3$ up to $T=75$~K. 
If a subsystem does indeed exist which does not undergo full
magnetic ordering then the magnetic entropy would take a nonzero
value below the transitions. Assuming that all of $S_{\mathrm{mag}}$
is accounted for by $T=75$~K, then the amount of magnetic entropy
associated with the partially ordered subsystem would be 
43.5~\% of $2 R \ln 3$. Such a value is possible
given that the amplitude $A_{2}$
associated
with the Gaussian component approximately twice $A_{1}$ at low
temperatures (Fig.\ref{fit}(a)) suggesting that around
two thirds of the spins are associated with the partially
ordered subsystem.

Finally, it has been suggested\cite{crawford} that the unusual ordering behavior and smaller than
expected magnetic entropy may stem from the integer spin, which may also be the
case for the recently reported pyrochlore Y$_2$Ru$_2$O$_7$ \cite{y2ru2o7}.

\section{Conclusion}

In conclusion we have investigated magnetic transitions in
GeNi$_{2}$O$_{4}$ using implanted muons. We observe two separate
contributions to the measured spectra below the upper
magnetic transition at $T_{\mathrm{N1}}$ suggestive of two
different magnetic environments coexisting in the material.
Our data may be
interpreted in terms of two distinct magnetic subsystems that undergo
separate ordering transitions. The first subsystem, which
accounts for the larger portion of the sample, partially orders
at the higher transition temperature $T_{\mathrm{N1}}$ 
leaving significant spin disorder that
persists to the lowest measured temperature. This subsystem
coexists with a second subsystem that is fluctuating dynamically
down to the lower transition temperature $T_{\mathrm{N2}}$,
where this second subsystem undergoes full magnetic ordering.
The exchange in amplitudes of the signals
corresponding to each subsystem suggests that the number of
Ni$^{2+}$ spins associated with each subsystem is not constant
but varies above $T \approx 9$~K. Further magnetic neutron
diffraction measurements would be required to elucidate the precise
nature of the proposed subsystems.

\acknowledgments 

Part of this work was carried out at the Swiss Muon Source, 
Paul Scherrer Institute, Villigen, Switzerland.
We thank Hubertus Luetkens and Robert Scheuermann for technical assistance
and Paul Goddard and Michael Brooks for useful discussion.
This work is supported by the EPSRC. 
T.L.\ acknowledges support from the European Commission under the 6th 
Framework Programme through the Key Action: Strengthening the European 
Research Area, Research Infrastructures. Contract no: RII3-CT-2003-505925


\begin{thebibliography}{99}
\bibitem{hcnote}
We note that the peaks observed in our ZF measurements appear slightly broader
than those observed previously, \cite{crawford,hara} but this
is the effect of a lower sampling frequency rather than an intrinsic
 property of our sample.
\bibitem{fits}
In all of the fits, the initial asymmetry $A(t=0)$ and the background
$A_{\mathrm{bg}}$ was held constant for all
temperatures. The relaxation
rate $\lambda_{1}$ was found also approximately constant in phase I, and was
fixed at 11.7~MHz. This resulted in three degrees of freedom that were
found to vary with $T$ acorss the measured temperature range.
\bibitem{note}
Beamtime was only available to study one sample of
GeNi$_{2}$O$_{4}$. In view of the fact that out x-ray and heat
capacity data are identical with those of previously published
work, we conclude that our $\mu^{+}$SR data are revealing of intrinsic
GeNi$_{2}$O$_{4}$ behaviour.
\end{thebibliography}
\end{document}